\documentclass[aps,pre,nofootinbib,showpacs,twocolumn]{revtex4}
\usepackage{amssymb,latexsym,mathrsfs}
\usepackage{amsfonts}
\usepackage{amsmath}
\usepackage{graphicx}

\newcommand{\beq}{\begin{equation}}
\newcommand{\eeq}{\end{equation}}
\newcommand{\beqn}{\begin{eqnarray}}
\newcommand{\eeqn}{\end{eqnarray}}
\newcommand{\bearr}{\begin{array}}
\newcommand{\enarr}{\end{array}}

\newcommand{\ra}{\rangle}
\newcommand{\la}{\langle}

\def\bea{\begin{eqnarray}}
\def\eea{\end{eqnarray}}
\def\ba{\begin{array}}
\def\ea{\end{array}}

\begin{document}
\title{Phase separation transition in  anti-ferromagnetically interacting particle
systems}
\author{ Anasuya Kundu}
\email[E-mail address: ]{anasuya.kundu@saha.ac.in}
\author{P. K. Mohanty}
\email[E-mail address: ]{pk.mohanty@saha.ac.in}
\affiliation{Theoretical Condensed Matter Physics Division,\\ Saha Institute of Nuclear Physics,
1/AF Bidhan Nagar, Kolkata, 700064 India.}
\date{\today}
\vskip 2.cm
\begin{abstract}
One dimensional non-equilibrium systems with short-range interaction can undergo
phase transitions from 
homogeneous states to phase separated states as interaction ($\epsilon$) among
particles is increased.  One of the model systems  where such  transition has been observed  is
the extended Katz-Lebowitz-Spohn (KLS) model with ferro-magnetically interacting particles
at $\epsilon=4/5$. Here, the system remains homogeneous for small interaction strength
($\epsilon<4/5$), and for anti-feromagnetic interactions ($\epsilon<0$). We show that the phase
separation transitions  can  also  occur in anti-ferromagnetic systems if interaction among
particles depends explicitly on the  size of the block ($n$) they belong to. We study this
transition in detail for a specific case $\epsilon = \delta/n$, where phase separation occurs for 
$\delta < -1$.
\end{abstract}
\pacs{xxx xxx }
\maketitle

One dimensional driven diffusive systems~\cite{dds,dds1} have been a subject of extensive studies
in recent years. Some driven diffusive models with local 
dynamics exhibit exotic phenomena like phase separation and phase
transitions~\cite{KLS,AHR,Lahiri,eKLS,nKLS,ABC},  
with novel spatial correlations in their steady states,  which
can not be seen for systems in thermal equilibrium, with short range interactions. 
Since there is no general theoretical method to study non-equilibrium systems,
phase transitions for them can not be inferred from any guiding principle. Few years
ago  a general criterion for the existence of phase separation in density-conserving
one dimensional driven diffusive systems, have been proposed by using a mapping of
these models to a Zero Range Process (ZRP) by Kafri {\it et al.}~\cite{criteria}.
It has been suggested  there, that existence of phase separation in any given model 
depends only on the rates (or the steady state curret $J_n$) at which  domains of 
various sizes ($n$) exchange particles.  A strong phase separation, where density fluctuations 
are limited to the domain boundaries,  occurs  for arbitrarily small densities 
when  the steady state current $J_n$ flowing through a block of size $n$ vanishes  for 
thermodynamically large blocks.  Whereas a  condensed phase with large  density  
fluctuations occurs when the current
\beq
  J_n \sim J_\infty (1+ b/n^\sigma) \label{eq:criteria}
\eeq  
with either $\sigma<1$  and $b>0$ or with  $\sigma=1$  and $b>2$.  In this case  
phase separation occurs only for large enough densities. 

\begin{figure}
\noindent \includegraphics[clip,width= 6cm, angle = 0]{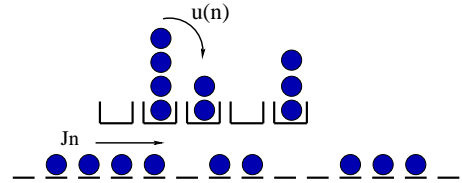}
\caption{\label{fig:zrp}
A microscopic configuration of driven diffusive system is mapped to corresponding
configuration in ZRP.}
\end{figure}

  This  criteria~\cite{criteria} is based on a mapping 
which relates driven diffusive systems with  a well known 
model  called Zero Range Process.  
In ZRP, each site on a lattice (formally called boxes) can accommodate more than one particle,
which are allowed to hop to the neighboring box with a rate $u(n)$ that depends on the number of
particles $n$ present in the departure box (see Fig~\ref{fig:zrp}).
The  vacant sites  of any particular  driven 
diffusive system  can be considered as the boxes and the domain of size $n$ 
to the immediate left of this vacant site can be considered as the number of 
particles. Finally, the steady state current $J_n$ can be regarded as the 
hop rate  $u(n)$ of ZRP. Such a mapping  can be shown to be exact~\cite{criteria}
in some models. In others~\cite{eKLS, nKLS}, though the ZRP mapping is only 
approximate, phase separation could be predicted beyond doubts.
Phase separation transition  with $\sigma=1,~ b>2$ was first observed in  a 
class of models~\cite{eKLS} where two species of hard-core particles (equal in number) on a one-dimensional 
ring interact ferro-magnetically, which was extended later~\cite{nKLS} for unequal density of particles. 
In these cases, $b$ depends on the interaction strength  $\epsilon$ and condensation  occurs for 
large density of particles as $\epsilon$  is increased beyond a critical  value $\epsilon_c.$ 
For $\epsilon<\epsilon_c$, including the antiferromagnetic interaction $\epsilon<0$,  the 
system  remains homogeneous. 

In this article we introduce a model with two  species of  particles which interact 
antiferromagnetically, where the interaction strength depends on the size of the domain.  
We show that such modifications can lead to phase separation transition, even for  anti-ferromagnetically 
interacting particles. An intuitive and physical argument is provided in favour of  
the possibility of such a transition.

The model is defined on a ring   where sites are labeled by $i=1,2,...L$.  Each site 
$i$ can either be vacant $(s_i=0)$ or occupied by a positive $(s_i=+1)$ or a negative 
$(s_i=-1)$ particle.  These particles interact  anti-ferromagnetically with  interaction 
strength  $\epsilon(n_l)$ which depends on the size $n_l$ of the $l^{th}$ block  that the particles 
belong. Note that a block (or a {\it domain}) here is  defined by an uninterrupted sequence of particles and 
leveled by an index $l=1,2\dots N_0$, where $N_0$ is the number of  vacancies in the ring. 
Formally the interaction can be represented by
\begin{equation}
H= -\frac{1}{4} \sum_{l=1}^{N_0}  \epsilon(n_l) \sum_{i\in n_l} s_i s_{i+1} .
\end{equation}
The model evolves according to the nearest-neighbor exchange rates
\bea
+-\, \mathop{\longrightarrow}^{1+\Delta H} \, -+; \hspace*{0.8cm}
+0\, \mathop{\longrightarrow}^{\alpha} \, 0+; \hspace*{0.8cm} 0-\, \mathop{\longrightarrow}^{\alpha} \, -0 .
\label{eq:dyn}
\eea
Clearly, the dynamics is particle  conserving.  Thus the system  can be  characterized by 
the relative densities $\rho= \frac{N_++N_-}{N_0}$ and
$\eta =  \frac{N_+}{N_++N_-}$ where $N_\pm$ are the number of $+$ and $-$  particles 
respectively.

First let us consider the non-interacting case, $\epsilon= 0$. Here, the steady state 
weight $W(\{k_l\})$ of all the configurations  in grand canonical ensemble (GCE) 
can be written as~\cite{eKLS,nKLS}
\bea
 W(\{n_l\}) = \prod_{l=1}^{N_0} z^{n_l} Z_{n_l}, 
\eea
where $n_l$ is the number of particles that reside in the $l^{th}$ block and $Z_{n_l}$ is the 
sum of all microscopic configurations of a block of size $n_l$.  The fugacity $z$ is associated 
with both  ($\pm$) kinds of particles.  
Since the weight of the configurations  are  factorized   in terms of weight $z^{n_l}Z_{n_l}$ 
of individual blocks, one can draw an analogy  of the model with  a  ZRP having $N_0$ boxes  
and $(N_+ + N_-)$ particles where steady state in GCE has a product measure; the single box weight 
of ZRP is $f(n) = Z_n$.  Further, from the exact  
results of ZRP in grand canonical ensemble $f(n)= \prod_{k=1}^n u(k)^{-1}$,  one gets   
$u(n) = Z_{n-1}/Z_{n}$  (as $Z_n= \prod_{k=1}^n  Z_k/Z_{k-1}$). 
Since in the non interacting  system,   the current  through a block of size $n$ is 
$J_n= Z_{n-1}/Z_n$, one identifies $u(n) =J_n$.  Now, from the assymptotic behaviour~\cite{eKLS} of 
$J_n= \frac 1 4 (1+ \frac{3/2}{n})$ one  can conclude, using criteria mentioned in
Eq.~(\ref{eq:criteria}), that phase separation is not possible in  this  case.

The exact mapping of the model to ZRP does not hold for interacting system. 
However,   it has been argued~\cite{eKLS} 
that  the correlations  between neighbouring boxes can {\it still}  be neglected for sufficiently 
large $\alpha$  and  the  correspondence  $u(n)= J_n$   provides  a correct prediction of  
phase separtion  transition. Since the  blocks are considered  uncorrelated, one can calculate $J_n$ 
by modeling  a  single block of size $n$ where particles follow  dynamics~(\ref{eq:dyn}). 
Steady state  weights of  such a system on a ring~\cite{KLS} has  an Ising measure 
$P(\{s_i=\pm\}) ~ \exp(-\beta H )$ with  
\begin{equation}
H=-\frac{\epsilon}{4}\sum_i s_i s_{i+1}+h\sum_i s_i, \label{eq:Ising}
\end{equation}
{\it i.e.,}  weight of  any configuration  is same as that of the above system in  equilibrium. 
The  second term is  needed to fix  magnetization of the Ising system  ${\cal M} = 2 \eta -1$. 
For equal density case,   $h=0$.    Thus,  the current 
\bea J_n &=&\la ++-+\ra_n   +   (1-\epsilon) \la ++--\ra_n \cr &+&  \la -+--\ra_n 
 +(1+\epsilon) \la -+--\ra_n    \label{eq:static}\eea   is a static correlation function of 
the Ising system Eq.~(\ref{eq:Ising}), which  can be calculated~\cite{Hager}  using   the standard transfer 
matrix formalism~\cite{baxter}.  For a  thermodynamically large block, current $J_\infty$  
calculated on a ring is  identical to that of an open domain. 
However, the corrections $J_n= J_\infty( 1+ b(\epsilon)/n)$  depends on the boundary conditions. 
It has been  argued~\cite{Krug} that  $b(\epsilon)$  for an open system is related to that of the  
ring by a universal factor;   $ b(\epsilon) = \frac 3 2 b_R$, where  the subscript $R$ stands for ring.  
Thus it is sufficient to  calculate   correlation functions Eq.~(\ref{eq:static}) on a ring with respect 
to  the equilibrium Ising measure. A  simple transfer matrix formalism reveals, 
\bea b(\epsilon,\eta=1/2)=  \frac{3}{4}  \frac{(2+\epsilon) \gamma +2\epsilon}{\epsilon+\gamma} ; \;\; 
 \gamma = \sqrt{\frac{1+\epsilon}{1-\epsilon}}+1. \eea 
  
\begin{figure}[h]
\noindent \includegraphics[clip,width= 6cm]{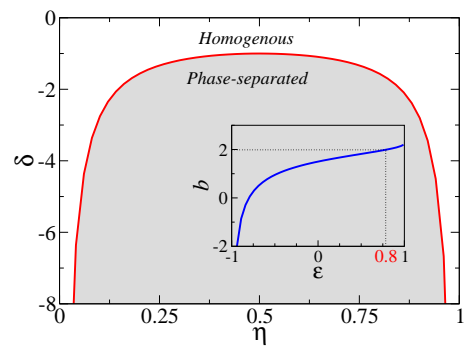}
\caption{  Phase diagram in $\eta$-$\delta$ plane: the shaded region corresponds to 
phase separated state.  Inset, here, shows  dependence of $b$  on $\epsilon$ for 
equal density case.} \label{fig:beps} 
\end{figure}

These relations predict that, for any relative density $\eta$ the system   undergoes a 
phase transition  from a homogeneous state to a phase separated state when $\epsilon$ is 
increased  beyond a critical threshold $\epsilon_c$.  
Inset of Fig.~\ref{fig:beps}  shows variation of $b(\epsilon)$   with $\epsilon$ for equal density 
case  $\eta=1/2$; here  $\epsilon_c= 4/5$.   It is evident from the figure, that   
phase separation is not possible  when $\epsilon<0$ (anti-ferromagnetic interaction). 
These  predictions for ferro-magnetically interacting system have been 
verified  using Monte Carlo  simulations of extented KLS (EKLS)
model for both $\eta=1/2$~\cite{eKLS} and for $\eta\ne 1/2$~\cite{nKLS}.

A possible reason, why antiferromagnetically  interacting particles  do not phase separate 
is the following. For ferro-magnetic case,  each particle likes the same kind of particles as 
neighbours and the asymmetric exchange $+- \to -+$, then arrange  more $+$ particles towards 
the right end of the block. With this  arrangement of particles, $0$s  enter  easily into 
the ferromagnetic block from both sides allowing the movement of  the whole block as a 
single entity. However, in an antiferromagnetic system,  $+$ and $-$ particles  prefer to 
be arranged alternatively which  restrict  $0$s from the left (right) to move further 
once it encounters a $+$ $(-)$ particle.
The antiferromagnetic interaction could compete with the boundary dynamics when 
$\epsilon  \sim {\cal O} (1/n)$, $[i.e. ~n\epsilon \sim {\cal O} (\alpha)]$. 
Thus,  it is suggestive that a phase separation could be possible  when  
$\epsilon_n= \frac{\delta}{n}$, explicitly depends on  $n$.
In the rest of the article we concentrate on  antiferromagnetic case with 
interaction strength $\epsilon_n= \frac{\delta}{n}$, that depends  explicitly 
on the size of the block. It is not difficult to  verify that the steady 
state  have Ising measure.  Thus,  $J_n$ can be calculated in a straightforward 
way following the  procedure discussed  here.   This results in,
 \bea
J_n &=&J_\infty\left(1+\frac{3-4 \delta J_\infty}{2n}\right);  
J_\infty= \eta(1-\eta) .\label{eq:Jdelta}
\eea    
Thus, $b(\delta, \eta) =(3-4 \delta J_\infty)/2$ and for any given $\eta$ 
phase separation occurs  for  systems having large density $\rho$ when 
$\delta < - [4 \eta (1-\eta)]^{-1}$. Corresponding phase diagram in $\eta$-$\delta$  
plane is shown in Fig. \ref{fig:beps}.  

\begin{figure}[h]
\noindent \includegraphics[clip,width= 6cm]{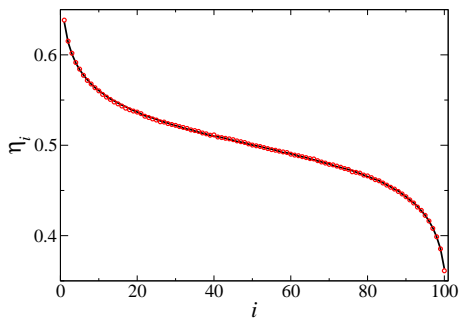}
\caption{\label{fig:prof}
Comparison of density profiles: solid line corresponds to the density profile of
a block of size $n = 100$ and $L = 500$.
Symbols correspond to that of a ring with $100$ particles and a single
defect $0$. In both cases, $\delta =-0.9$.}
\end{figure}

To  check the correctness of the prediction, we choose to study
only the equal density case in details and verify  
the predictions there.  In this case,  $J_\infty=1/4$ and 
$b(\delta)=\frac{3-\delta}{2}$; thus the transition is expected for 
$\delta<-1$.  Note, that  in deriving Eq.~(\ref{eq:Jdelta}) we have assumed that,  
a domain containing $n$ particles of any species, behaves like an open
system where $(+)$ particles enter from left and $(-)$ particles exit from right
with rate $\alpha$. Hence, the dynamics of a domain can be modelled by a ring having 
$n$ particles $(\pm)$ and a single defect $0$, and following above dynamics. To check 
the validity of this assumption we have 
calculated the  density profile  of a block of size $n=100$   using 
Monte Carlo simulation  of  a system of size $L=500$  at $\delta =-0.9$.  
In Fig.~\ref{fig:prof}  the
density profiles of a block of size $n=100$  obtained from Monte Carlo simulation of  
the full system is compared with  that  obtained from  a ring with $100$ particles and a 
defect $0$. In both cases we choose $\delta=-0.9$.  An excellent match seen here,
justifies the assumption that a domain of size $n$ in fact behaves like a ring having
$n$ particles and a single defect~$0$. 

\begin{figure}[h]
\noindent \includegraphics[clip,width= 6cm]{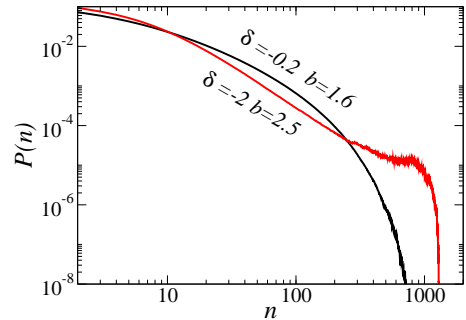}
\caption{\label{fig:pk}  Block size distribution for $\delta=-0.2$ (black)
and $-2.0$ (red online) for a system with $L=2000$ and $N_0=100$.   }
\end{figure}

The  coarsening dynamics of EKLS model is quite slow. Thus in  fact achieving  a true 
condensate   through  Monte Carlo simulations, in a finite time, is  practically impossible. 
However, the distribution of block sizes $P(n)$,  provides an indication: 
when $\delta<\delta_c=-1$,  a condensate appears along with the  usual scale-free distribution. 
In fact, to observe the condensate  one must take the density much larger than 
$\rho_c= 1/ (b(\delta)-2)$. 
Figure~\ref{fig:pk} shows distribution of size of the block for $\delta=-0.2$ and
$-2.0$, correspondingly the values of $b$ are $b=1.6$ and $2.5$. The relative density is given by
$\rho=19$ with $L=2000$. 
It is clear that condensation emerges only in the latter case. 
 
\begin{figure}[h]
\noindent \includegraphics[clip,width= 9cm,height=5 cm]{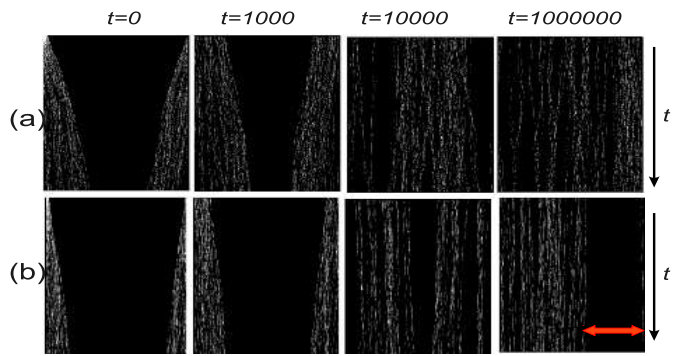}
\caption{Evolution of a phase separated configuration: the snapshots of the system   
are taken at $t=0, 10^3, 10^4,$ and $10^6$. (a) Upper and 
(b) lower pannels  correspond to $\delta =-0.2$ and $-2.0$ respectively. 
Here $L=2000$ and $N_0=100$. The double arrow (red) corresponds to a large condensate that
survives.}\label{fig:evol}
\end{figure}

It is expected that, in the phase-separated regime, any initial phase-separated 
configuration evolves only to generate fluctuation in the size of the condensate; the 
condensate does not disappear. To check the stability of an initially phase separated configuration 
we have done Monte Carlo simulation of the system of size $L=2000$ with $N_0=100$ and 
look at the  space-time plots for $2000$ MCS at $t=0,10^3,10^4$ and $10^6$. The snapshots for 
$\delta =-0.2$ and $-2.0$ are shown in Fig.~\ref{fig:evol}. Clearly for the latter case, a large 
condensate (indicated by a double arrow) survives even at $t=10^6$.

In conclusion, we have studied the  extended KLS model  with antiferromagnetically  interacting particles. 
The strength of interaction between particles $\epsilon(n)= \delta/n$ in this  model depends explicitly on the size of 
the block $n$ they belong to.  The model  can be mapped to zero range process by identifying vacancies as boxes; the hop 
rate $u(n)$ in ZRP  is identified as the   particle current $J_n$ in the lattice, which  can be calculated 
exactly. Such a  mapping enables one to predict that the phase separation transition
(same as condensation transition in ZRP) occurs for $\delta<-1$.
These predictions are verified by using Monte Carlo simulations.  

Some comments are in order.  Unlike the particle current  in the lattice, the hop rate in ZRP is  
stochastic  and  uncorrelated in time.  Recent studies~\cite{tZRP}  of ZRP  with time-correlated hop rate 
suggest that  effective $b$ which  sets the criteria for condensation transition may   change when the correlation 
is  significantly large.  However, our  numerical study   of the antiferromagnetic system indicates
that the time-correlations, present in $J_n$, are possibly irrelevant. In fact, such
correlations are known~\cite{criteria} to be irrelevant in AHR~\cite{AHR} model.

\end{document}